\documentclass[final,3p,times,onecolumn]{elsarticle}

 \usepackage{graphicx}
\usepackage{epsfig}
\usepackage{amssymb}
 \usepackage{amsthm}
\usepackage{color}

\journal{Journal of Non-Newtonian Fluid Mechanics}
\begin{document}
\begin{frontmatter}
\title{Comparative investigations of surface instabilities ("sharkskin") of a linear and a long-chain branched polyethylene}
\author{Teodor I. Burghelea, Hans J\"{u}rgen Griess, Helmut M\"{u}nstedt}

 \address{Institute of Polymer Materials, Friedrich-Alexander-University Erlangen-N\"{u}rnberg, D-91058 Erlangen, Germany}
\begin{abstract}
An experimental study of the physical origin and the mechanisms of the sharkskin instability is presented. Extrusion flows through a slit die are studied for two materials: a linear low density polyethylene (LLDPE) which exhibits sharkskin instability for flow rates larger than an onset value and a low density polyethylene (LDPE) which does not show any instability over a broad range of flow rates. 
By combining laser-Doppler velocimetry (LDV) with rheological measurements in both uniaxial extension and shear, the distributions of tensile and shear stresses in extrusion flows are measured for both materials. The experimentally measured flow fields appear to be qualitatively similar for both the unstable (LLDPE)  and stable case (LDPE): around the die exit the flow accelerates near the boundaries and decelerates around the flow axis.  The fields of the axial gradients of the axial velocity component are, however, quite different in the two cases. In the unstable case there exists a strongly non-uniform transversal distribution of velocity gradients near the die exit. This non-uniformity of the distribution of gradients is significantly smaller in the stable case. Significant differences in the extensional rheological properties of the two materials are found as well. Due to its branched structure, the LDPE is able to sustain higher tensile stresses prior to failure. Measurements of the distributions of tensile stresses around the die exit reveal a stress boundary layer and a stress imbalance between the boundaries and the bulk. The magnitude of the stress imbalance exceeds the melt strength in the experiments with the LLDPE which causes the failure of the material in the superficial layers and results in the emergence of the sharkskin instability. In the experiments with the LDPE, the magnitude of the stress imbalance remains smaller than the melt strength which explains the lack of an instability. The measured shear stresses around the die exit are significantly smaller than the tensile stresses, suggesting that the shear component of the flow plays no significant role in the emergence of the sharkskin instability.
\end{abstract}

\end{frontmatter}

\newpage


\newpage

\section{Introduction} \label{sec:Introduction}
Understanding and reducing polymer instabilities are of enormous economical importance, as they are limiting factors of extrusion efficiency. 
Flow instabilities during the extrusion of polymer melts have been observed since nearly six decades ago and sharkskin instabilities, in particular, were first reported in late 60's, \cite{howells, tordella}.  
Such instabilities have been extensively reviewed during the past three decades \cite{petrie, larson, dennreview}. In spite of a large amount of both experimental and theoretical studies of the sharkskin, the origins and the physical mechanisms of this phenomenon are not yet fully understood. It has been agreed, however, that the sharkskin instability is most probably related to both the flow kinematics and the local dynamics near the exit of the die.   

\begin{figure*}[h]
\begin{center}
\centering
\includegraphics[width=9cm]{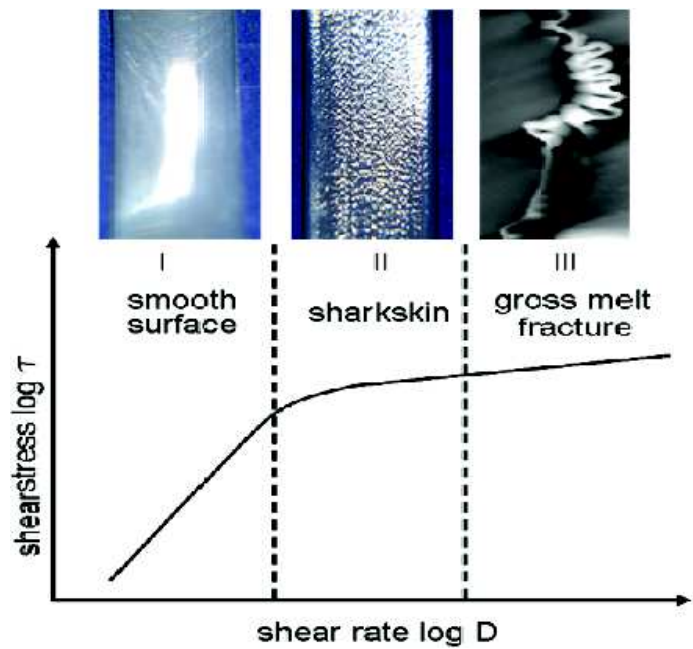}
\caption{Examples of different types of extrusion instabilities observed for a linear polyethylene \cite{merten} at different values of the apparent shear rate and the wall shear stress.} \label{fig_phasediagram}
\end{center}
\end{figure*}

Fig. \ref{fig_phasediagram} shows a "cartoon" version of the stability diagram experimentally measured by Kalika and Denn \cite{denn2} which illustrates various states of the extruded surface observed for a linear polyethylene as a function of apparent shear rate and wall stress.
 At low apparent shear rates the extrudate is smooth (regime I). Especially linear polymers show a transition to a surface instability at relatively low shear rates that is referred to as  "sharkskin" (regime II). Sharkskin is characterized by small amplitude (and high frequency) surface distortions  \cite{denn1}. 
 A first indication of the onset of sharkskin is a loss of the gloss at the polymer surface \cite{ramamurthy}. Previous investigations and our own results indicate that surface instabilities occur above certain wall shear stresses. In the case of LLDPE the severity of surface melt fracture increases under apparently steady flow conditions with increasing shear rates or stresses. At still higher shear rates gross melt fracture occurs (regime III).
It is believed that sharkskin is caused by a localized stress concentration at the die exit. Both, numerical simulations and flow birefringence experiments refer to a stress concentration at the die exit \cite{arda}. One assumption that explains the origin of sharkskin implies the existence of high tensile stresses at the die exit \cite{cogswell} which lead to an extensional failure of the material. Particularly, the surface layers of materials that stick to the wall within the die are highly accelerated after exiting it. The acceleration causes an extension of the material in the flow direction which results in tensile stresses. In the case the tensile stress exceeds the melt strength the material breaks and forms a crack on the surface near the die exit, which leads to a momentary reduction of the tensile stress to a value that the material can withstand before the stress escalates again. This periodic process causes the phenomenon of sharkskin. The above explanation has been supported by El Kissi and Piau \cite{Piau}, who visually observed cracks perpendicular to the flow direction during the extrusion. Venet and Vergnes relate the emergence of sharkskin to the stress levels along local streamlines close to the extrudate surface, \cite{venet1, venet2}. Rutgers and Mackley \cite{rutgers} proposed that a rupture mechanism could explain the correlation between sharkskin and melt strength. 

Though there exists little doubt that the sharkskin is indeed caused by an extensional failure of the material near the die exit \cite{denn3},  a rigorous quantitative description of the phenomenon is missing. This is mostly due to the fact that a complete theoretical picture of different mechanisms of failure of polymeric materials remains an elusive goal. The simple scaling theory of failure in the fast stretching regime proposed by Joshi and Denn \cite{denn4} is, most probably, unable to explain the sharkskin because the emergence of sharkskin implies a reorganization of the free surface of the material accompanied by a stress release.

The present study explores the physical origins of the sharkskin phenomenon by combined measurements of the flow kinematics (flow fields and fields of axial velocity gradients) and measurements of the rheological properties of the materials in both uniaxial extension and shear.  The distributions of tensile and shear stresses in the flow are obtained by merging the kinematic data with the rheological data.  This method has several advantages over classical birefringence measurements \cite{arda}:
\begin{enumerate}
\item It can measure separately the tensile and shear stresses.
\item It can resolve the distributions of both tensile and shear stresses at the corners regions near the die exit.
\item It provides quantitative stress measurements which can be directly compared with the melt strength measured by an extensional rheometer.
\end{enumerate}
To gain additional insights into the emergence of the sharkskin instability, a comparative analysis of the stress distributions corresponding to both an unstable (sharkskin) and stable extrusion flows is performed.

\section{Outline} \label{sec_organization}
The paper is organized as follows. The experimental setup, the measuring techniques, the physical properties of the polymer melts and the analysis procedure of the flow data are discussed in Sec. \ref{sec:Devices}.  A detailed description of the flow kinematics is given in Sec. \ref{subsec:kinematics}. A comparative analysis of the rheological properties of the materials in uniaxial extension is performed in Sec. \ref{subsec:rheology}. Measurements of the distributions of tensile stresses in the flow are presented and discussed in Sec. \ref{subsec:extensionalstress}. The role of the shear stresses in the emergence of the sharkskin instability is discussed in Sec. \ref{subsec:shearstress}.
The paper closes with a concise discussion of the main findings, Sec. \ref{Conclusions}. 

\section{Experimental section} \label{sec:Devices}
\subsection{Experimental setup and measuring techniques}\label{subsecsec_setup}

\begin{figure*}
\begin{center}
\centering
\includegraphics[width=10cm]{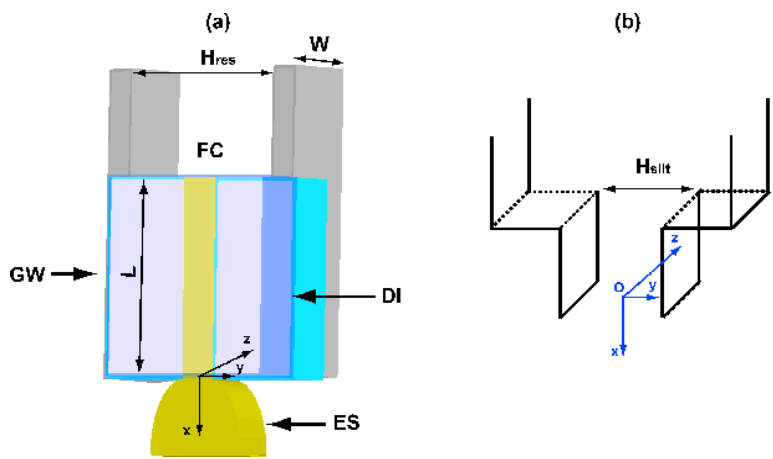}
\caption{(a) Schematic view of the experimental setup: \textbf{FC} - flow channel, \textbf{DI} -die inserts, \textbf{GW} - glass window, \textbf{ES} - extruded strand (b) Coordinate system.}\label{fig_setup}
\end{center}
\end{figure*}

The experimental setup is schematically presented in Fig. \ref{fig_setup} (a). It consists of a slit die with a rectangular cross-section. 
The dimensions of the slit die are $50 \times 14 \times 1 mm$ $(L \times W \times H_{slit})$. For all measurements the entrance angle of the slit die was $180~\symbol{23}$. The origin of the coordinate system was chosen in the plane of the die inlet at a distance of $7~mm$ from the glass window, Fig. \ref{fig_setup} (b). A pressure transducer in the slit allow the determination of the wall shear stress. The temperature was controlled within $\pm~0.2~\symbol{23}C$ along the flow channel and over the duration of an experiment.

A commercial two component laser-Doppler velocimeter from Dantec Dynamics was used to investigate the flow fields of the two polyethylenes. 

Two pairs of laser beams are focused in the flow channel through the transparent glass window of the setup and the local flow velocity is assessed by measurements of the Doppler shift of the light scattered by small seeding particles added to the polymer melt,  \cite{ldabook}. $TiO_2$ particles with sizes ranging from $0.1~\mu m$ to $1~\mu m$ have been used as seeding particles.
The LDA probe is mounted on a three dimensional translational stage (from Dantec Dynamics) with a resolution of $10 \mu~m $ on each axis.
The flow speed can be reliably measured down to $250~ \mu m/s$. Within this velocity range the accuracy of the velocity measurements is about $10~\%$ and becomes better than $1~\%$ with increasing velocities. A more detailed description of the LDV system and the mechanical setup for generating a constant throughput of the melt are given in \cite{schmidt}. The coordinate system used are shown in Fig. \ref{fig_setup} (b). 

\subsection{Molecular and physical properties of the materials investigated} \label{sec:Characterization}
The materials selected in this study are a linear low-density polyethylene (LLDPE) and a low-density polyethylene (LDPE).
The LLDPE used is a commercial product from Exxon called Escorene LLN 1201 XV which is mainly used for the manufacturing of films.
A detailed investigation of the molecular and rheological properties of the LDPE has been recently reported, \cite{utte}.    
 The LDPE used was Lupolen 1840 H from Basell, a commercial product frequently used for film extrusion and film blowing. Its molecular and rheological properties have been extensively investigated by Hertel, \cite{hertel}. It has random-like long-chain branched molecular structure which originates in its radical polymerization at high pressures, as opposed to LLDPE which has a linear molecular structure.
Characteristic material data are summarized in Table \ref{tab_tab1}.  Due to substantial differences in their molecular structure, the materials have different melt temperature $T_m$, different molar mass $M_w$ and different distributions $M_w/M_n$ of the molar mass, Table \ref{tab_tab1}. 

\begin{table*}
\begin{center}
  \begin{tabular}{@{} |c|c|c| @{}}
    \hline
   ~ & $\mathbf{LDPE}$ & $\mathbf{LLDPE}$ \\    \hline \hline
    $\mathbf{\rho_{25^{\circ} C}(g~cm^{-3})}$ & 0.919 & 0.926 \\ \hline
    $\mathbf{T_m(^{\circ} C)}$ & 110 & 125 \\  \hline
    $\mathbf{M_w(kg~kmol^{-1})}$ & 256 & 148 \\ \hline
    $\mathbf{M_w/M_n}$ & 15 & 4.2 \\ 
    \hline
   \end{tabular}
   \end{center}
 \caption{Physical properties of the polyethylenes investigated.} \label{tab_tab1}
\end{table*}

The choice of the two polyethylenes was motivated by the fact that the LLDPE shows sharkskin under the experimental conditions and LDPE does not show sharkskin and thus a comparative study is possible. Both materials were investigated with regard to differences in flow behavior and melt strength with the aim to explain their different surface structures. 

\subsection{Rheological characterization of the materials}\label{subsec_rheochar}

As it will be shown in detail through the paper, the flow of the polymer melts through the die combines both shear and biaxial extension and therefore a characterization of the stress distribution in the flow requires the knowledge of the rheological properties of the melts in both extension and shear. Measuring the extensional rheological response of the material under such kinematic conditions was not possible and therefore we have investigated the transient response of the materials in uniaxial elongation, only.  
The rheological characterization of the materials in uniaxial extension has been done using a M\"{u}nstedt type extensional rheometer built in the house.  A detailed description of this device and its operating principle can be found elsewhere, \cite{muenstedt}. 
The elongational specimens were manufactured by extrusion through a capillary followed by a retardation process in silicone oil. The samples were glued to aluminum clamps. For the elongational tests, specimens with a length of $20~mm$ and a diameter of approximately $3~mm$ were used. 
The specimen under investigation is clamped between the plates of the rheometer and immersed in a silicone oil bath to minimize gravity and buoyancy effects. A secondary role of the oil bath is to ensure isothermal conditions during the extensional tests. While the bottom plate of the rheometer is stationary, the top plate of the rheometer is moved vertically by an AC-servo motor controlled by an analogue to digital converter installed on a personal computer. The force acting on the bottom plate and the position of the top plate (which defines the actual Hencky strain) are transferred in real time to a personal computer by an analogue to digital card. Elongational measurements are performed at constant rate of deformation $ \dot \epsilon$. The maximal value of the rate of deformation accessible by the rheometer was $3s^{-1}$. This upper bound could be increased only by using the time-temperature superposition principle.  

The rheological characterization of the materials in shear has been done using a stress controlled shear rheometer ($AR-G2$ from TA Instruments) equipped with a magnetic bearing.

\subsection{Analysis of the flow fields}\label{subsec_flowfields}

Velocity profiles were measured at various positions within the die ($x \in [-50, 0]~(mm)$) as well as outside the die ($x \in [0, 1.5]~(mm)$). Whereas inside the die the axial distance between consecutive profiles was $\Delta x = 5~mm$, around the die exit it has been reduced down to $\Delta x = 0.1~mm$. This allows one to characterize the flow fields near the die exit with increased spatial resolution.
As a next step, the velocity profiles are fitted by the power law below, as suggested in \cite{schmidt}.
\begin{equation}\label{equ:1}
v_{x}=v_{0} \left [ 1-\left | \frac{2x}{H} \right |^ {\frac{n+1}{n}} \right ]
\end{equation}
$ v_{0} $ is the maximum of the velocity and $n$ the power law index of the material. The power law fit allows both an extrapolation of the measured points to the die wall and a noise free calculation of the spatial derivatives of the flow fields. As the velocity profiles acquired within the die seem to systematically extrapolate to zero at the die walls, it can be concluded that there exists no wall slip for the materials we have investigated. The two dimensional fitted velocity data has been interpolated on an equally-spaced x-y grid which further allows calculation of the velocity gradients, $\frac{\partial V_x}{\partial x}$ and $\frac{\partial V_x}{\partial y}$.  
The two dimensional fields of velocity gradients are ultimately converted in two dimensional stress fields using both the elongational and the shear rheological data.

\begin{table*}
\begin{center}
  \begin{tabular}{@{} |c|c|c|c|c|c|c| @{}}
    \hline
 \textbf{ Experiment} & \textbf{ Material} & $\mathbf{T(^{\circ} C)}$ &  $\mathbf{\dot {m} ~(g~min^{-1}) }$ & $\mathbf{D~(s^{-1})}$ & $\mathbf{a_{xx} (mm~s^{-2})}$ & \textbf{sharkskin} \\    \hline   \hline
\textbf{1}  & LDPE & 135 & 1.4 & 12.5 & 10 & \textbf{No} \\  
 \hline
\textbf{2} &  LDPE & 180 & 9& 85 &510 & \textbf{No} \\   
  \hline
\textbf{3} &  LDPE & 180 & 14.4 & 135 &1170 & \textbf{No} \\      
  \hline 
\textbf{4} &  LDPE & 180 & 18& 168 & 2010 & \textbf{No} \\     
  \hline
\textbf{5} &  LDPE & 220 & 7.4& 71 & 265 & \textbf{No} \\     
  \hline
\textbf{6} &  LLDPE & 135 & 1.4 & 12.5 & 6.4 & \textbf{No} \\     
  \hline
\textbf{7} & LLDPE & 220 & 4.5 & 43 & 100 & \textbf{Yes} \\
  \hline
\textbf{8} &  LLDPE & 220 & 7.4 & 70 & 200 & \textbf{Yes} \\
  \hline
\textbf{9} &  LLDPE & 220 & 9 & 85 & 320 & \textbf{Yes} \\               
  \hline
   \end{tabular}
   \end{center}
 \caption{Investigations on sharkskin under different extrusion conditions. $\dot {m}$ is the mass throughput (flow rate), $D$ the apparent shear rate, and $a_{xx}$ the acceleration according to equation (2).
} \label{tab_tab2}
\end{table*}

Systematic measurements of the flow fields have been carried out for both LLDPE and LDPE at various flow rates (or apparent shear rates) and temperatures which are listed in Table \ref{tab_tab2}.

\section{Results and Discussion} \label{sec:Results}

\subsection{Kinematics of unstable (sharkskin) and stable (no sharkskin) extrusion flows}\label{subsec:kinematics}
Previous work has shown  that the sharkskin instability is initiated near the die exit, \cite{moynihan}.

\begin{figure*}[h]
\begin{center}
\centering
\includegraphics[width=8cm]{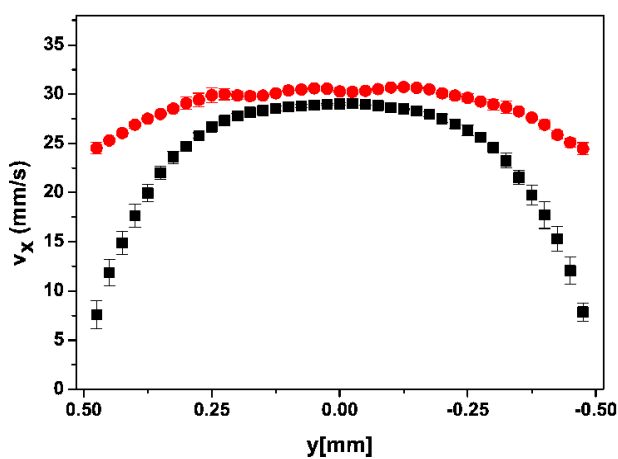}
\caption{Velocity profiles for a LLDPE with and without a fluoropolymer additive \cite{schwetz}: circles - LLDPE with fluoropolymer at $T=220 ^{\circ}~C$ , $D = 146~s^{-1}$, squares - LLDPE without fluoropolymer at $T=220 ^{\circ}~C$ , $D = 128~s^{-1}$. The data has been acquired inside the die, $x=-20 ~mm$.}\label{fig_fluo}
\end{center}
\end{figure*}

In a previous work \cite{schwetz} it has been demonstrated that sharkskin can be diminished or even suppressed, if the material slips at the die wall. This is consistent with the ideas of Cogswell \cite{cogswell} who suggested that the sharkskin is related to the abrupt changes of the flow conditions at the die exit. The acceleration decreases with the velocity difference $ \Delta v_{x} $ between the surface layers inside and outside the die. This fact indicates how the acceleration forces can be changed, namely by varying the flow behavior at the die wall. That can be done by inducing slip at the wall of the die which is a commonly applied measure in practice.
Fig. \ref{fig_fluo} illustrates a quantitative picture of such a modification. It exhibits the velocity profile of a pure LLDPE in comparison to that of a LLDPE with a fluoropolymer additive. The pure LLDPE sticks to the wall and the surface of the extruded strand displays distinct sharkskin. Upon addition of the fluoropolymer the material slips at the die wall, tensile stresses are no longer accumulated at the surface layers of the material around the die exit and, consequently, the sharkskin has disappeared. This result is in agreement with previous experimental observations \cite{piau1, wang} which show unequivocally that wall slip induced by the addition of fluorinated polymers to the melt suppresses the sharkskin instability.

An alternative and ingenious way of suppressing the sharkskin instability is to locally heat the die walls,  \cite{miller}. This way, the viscosity of the material is reduced in the surface layer and accumulation of the tensile stresses may be diminished up to the point the sharkskin instability disappears.

It has been previously demonstrated that the sharkskin observed for LLDPE is not caused by a change of the flow conditions within the die \cite{merten}. It is rather assumed that the accelerations of the surface layers of the exiting strand are the reason for the generation of sharkskin.
To get a deeper insight into the flow kinematics, measurements in the die exit region were carried out for both polyethylenes. Flow measurements with LLDPE were performed at several temperatures and values of the apparent shear rate $D$ (see Table \ref{tab_tab2}) both below and above the onset of the sharkskin instability. The apparent shear rate has been defined as $D= \frac{6 Q}{W H_{slit}^2}$ where $Q$ is the volumetric flow rate.  The measurements above the onset of the instability and right below it will be particularly discussed through this study.

We note that for both materials investigated the transversal velocity component $V_y$ was zero everywhere inside the die and nearly ten times smaller than $V_x$ around the die exit. For this reason we will focus in the following only on the fields of the axial component and the corresponding fields of gradients. The smallness of the transversal velocity component as compared to the axial component also suggests that one can neglect the extension of the material elements in the transversal direction $\frac{\partial V_y}{\partial y}$ and, consequently, approximate the flow around the die exit by an uniaxial extension.  

\begin{figure*}[h]
\begin{center}
\centering
\includegraphics[width=12cm]{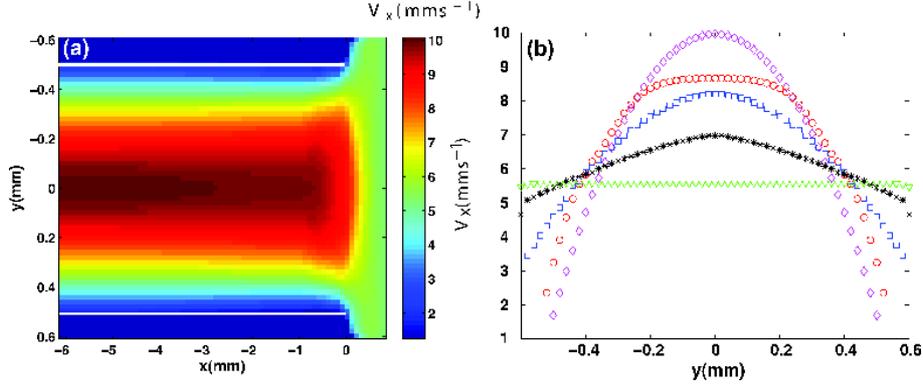}
\caption{(a) Spatial distribution of the axial velocity component $V_x$. The full lines indicate the location of the die walls.(b) Sample velocity profiles at several axial locations: diamonds - $x= -3 ~mm$, circles - $x=0~mm$, squares-$x=0.1~mm$, stars - $x=0.3~mm$, triangles - $x=0.6~mm$. The data refers to experiment $7$, Table \ref{tab_tab2}.} \label{flowmapLLDPE}
\end{center}
\end{figure*}

Fig. \ref{flowmapLLDPE} (b) presents several velocity profiles inside the die, at the die exit and outside the die. 
Inside the die ($x<0$) the velocity profiles can be described by Eq. \ref{equ:1}, do not depend on the axial coordinate $x$ and extrapolate to zero near the die walls (diamonds, Fig. \ref{flowmapLLDPE} (b)). 
At the die exit and outside of the die significant changes of the velocity profiles are visible (circles, squares, Fig. \ref{flowmapLLDPE} (b)). As one moves downstream ($x>0$), the flow speed decreases in the bulk region of the flow ($y \in [-0.4~0.4]~mm$) and increases in the boundary region of the flow.
It is obvious that especially the surface layers of the polymer become strongly accelerated over a very short distance from an initial velocity $ v_{x0} = 0 $ at the die exit to the final velocity $ v_{xs} $ of the strand. This acceleration may cause high forces in the surface layers which may ultimately lead to the failure of the material and the emergence of the sharkskin instability.
The acceleration $ a_{xx} $ of the surface layers can be estimated by \cite{merten1}.
\begin{equation} \label{equ:2} 
a_{xx}=\frac{\Delta v_{x}}{\Delta x}\left[v_{x0} +\frac{\Delta v_{x}}{2} \right] 
\end{equation}
where $\Delta v_{x}$ is the velocity difference between two profiles separated by the axial distance $ \Delta x $ and $ v_{x0} $ is the velocity of the surface layer of the extruded strand at the die exit.

As constitutive theories for polymeric materials involve velocity gradients rather than accelerations \cite{birdbook}, monitoring the acceleration of the surface layers might seem, at least at a first glance, somewhat unusual and deserves a brief discussion. 
\begin{figure*}[h]
\begin{center}
\centering
\includegraphics[width=13cm,height=5cm]{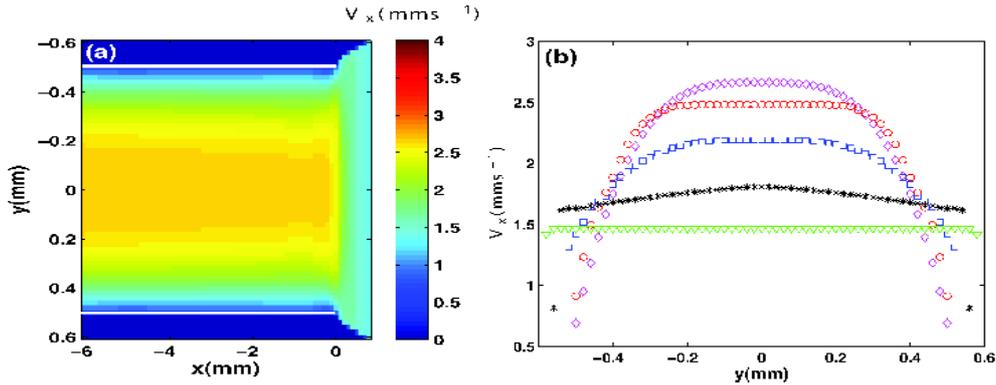}
\caption{(a) Spatial distribution of the axial velocity component $V_x$. The full lines indicate the location of the die walls. (b) Sample velocity profiles at several axial locations: diamonds - $x= -3 ~mm$, circles - $x=0~mm$, squares-$x=0.1~mm$, stars - $x=0.3~mm$, triangles - $x=0.6~mm$. The data refers to experiment $1$, Table \ref{tab_tab2}.} \label{flowmapLDPE}
\end{center}
\end{figure*}
Following ideas similar to the widely known Taylor hypothesis \cite{taylor1, taylor2}, one can quickly realize that over short distances the acceleration $a_{xx}$ is in fact proportional to the axial gradient of the axial velocity component, $\frac{\partial V_x}{\partial x}$. This is also visible in Eq. \ref{equ:2} if the second order term is assumed to be negligible over short distances. Therefore, accelerations would be interpreted through our manuscript as axial gradients of the axial velocity component or as local rates of extension.

The spatial distribution of the axial velocity component $V_x$ has been measured for LDPE, Fig. \ref{flowmapLDPE} (a). The velocity distribution for this stable (no sharkskin) case is qualitatively similar to that measured for the unstable case Fig. \ref{flowmapLLDPE} (a).
A flow deceleration in the bulk region and an acceleration in the boundary region are also observed for LDPE . 
For both LLDPE and LDPE, the extent of the flow fields along the y axis increases as the materials exit the die, indicating a significant die swell effect, Fig. \ref{flowmapLLDPE} (a), \ref{flowmapLDPE} (a).

This qualitative similarity of the flow fields between the sharkskin and the no sharkskin cases suggests that a proper understanding of the sharkskin phenomenon requires a deeper analysis of the flow kinematics and, perhaps, a correlation of the flow kinematics with the extensional rheological properties of the materials. 

In the following we focus on the spatial distribution of axial gradients of the axial velocity component $\frac{\partial V_x}{\partial x}$ which, as already pointed out above, provides an information on the distribution of accelerations as well.  

\begin{figure*}
\begin{center}
\centering
\includegraphics[width=12cm]{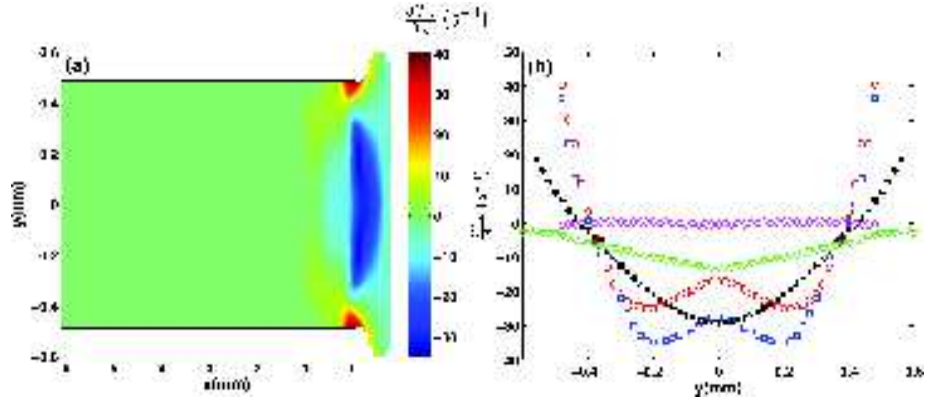}
\caption{(a) Spatial distribution of the rate of axial extension rate $\frac{\partial V_x}{\partial x}$. The full lines indicate the location of the die walls.(b) Sample profiles of the rate of axial extension at several axial locations downstream: diamonds - $x= -3 ~mm$, circles - $x=0~mm$, squares-$x=0.1~mm$, stars - $x=0.3~mm$, triangles - $x=0.6~mm$ .The data refers to experiment $7$, Table \ref{tab_tab2}.} 
\label{gradmapLLDPE}
\end{center}
\end{figure*}

Measurements of the axial velocity gradients for LLDPE above the onset of sharkskin are presented in Fig. \ref{gradmapLLDPE}. 
The rates of axial extension are practically zero within the die ($x< -1~ mm$), diamonds Fig. \ref{gradmapLLDPE}(b).  As one approaches the die exit ($x=0$) a clear spatial pattern of the axial gradients is visible, Fig. \ref{gradmapLLDPE}(a) .  The distribution of the axial gradients is symmetric with respect to the main flow direction (x), Fig. \ref{gradmapLLDPE}(b).

Large (and positive) values of the axial velocity gradients are visible near the corner region ($y= \pm 0.5~mm$) around the die exit which suggests that the readjustment of the flow from a no-slip to a free boundary condition is accompanied by a significant stretching of the material (or a high acceleration of the surface layers). The bulk of the flow is characterized by negative values of the axial gradients, consistent with a deceleration of the material elements. 

There are two important points to make regarding the transversal distribution of the gradients near the die exit (circles Fig. \ref{gradmapLLDPE}(b)). First, the absolute value of the gradients near the free surface of the flow is about twice as large as in the bulk. This suggests that the accumulation of extensional stresses is larger near the free surface of the material than in the bulk. Secondly, the transversal distribution of the gradients is strongly non-uniform (note the pronounced local maxima and the minimum of the profile around the die exit (circles and squares Fig. \ref{gradmapLLDPE}(b)) which suggests that, most likely, the extensional stresses are non uniformly distributed along the y axis as well. 

\begin{figure*}
\begin{center}
\centering
\includegraphics[width=12cm]{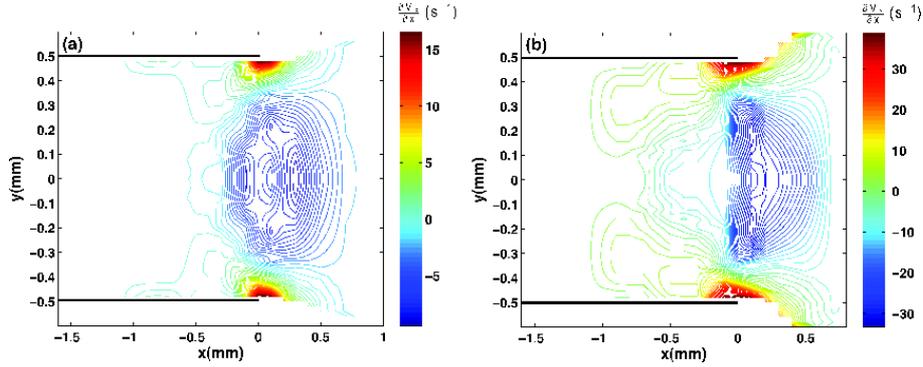}
\caption{Iso - contours of axial velocity gradients $\frac{\partial V_x}{\partial x}$ for LLDPE: (a) right below the onset of sharkskin instability (experiment  $6$, Table \ref{tab_tab2}) (b) above the onset of sharkskin (experiment  $7$, Table \ref{tab_tab2}) . The full horizontal lines indicate the position of the die walls.} \label{fig_comparisongradients}
\end{center}
\end{figure*}

To understand the origins of this particular distribution of the axial velocity gradients in the unstable case, we focus on the fields of axial velocity gradients right below the onset of sharkskin instability.  
Fig. \ref{fig_comparisongradients} presents a comparison of the topology of the fields of axial velocity gradients $\frac{\partial V_x}{\partial x}$ for LLDPE  right below the onset of the sharkskin instability (panel (a)) and above the onset of the sharkskin instability (panel (b)).

The iso-contours presented in Fig.  \ref{fig_comparisongradients} indicate that the particular distribution of velocity gradients in the flow is directly related to the transition to an unstable (sharkskin) regime. Indeed, all the features of the field above the onset of sharkskin instability  Fig.  \ref{fig_comparisongradients} (b) are already emerging right below the onset of the instability, Fig.  \ref{fig_comparisongradients} (a). We therefore conclude that there exists a clear relation between the particular distribution of the axial gradients in the flow and the emergence of the sharkskin instability.

\begin{figure*}
\begin{center}
\centering
\includegraphics[width=12cm]{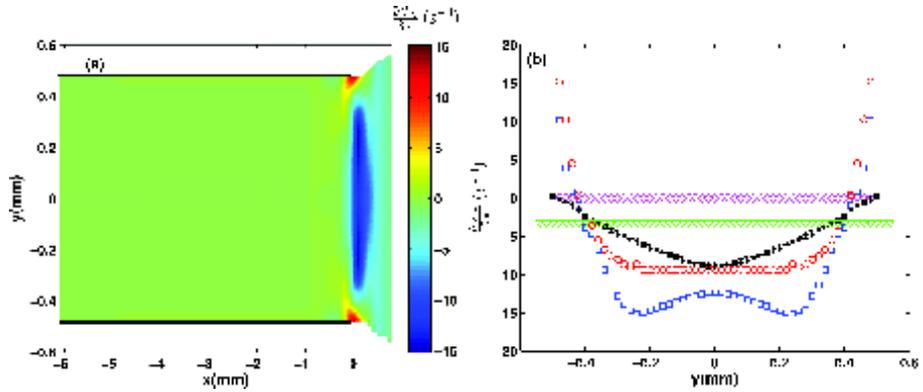}
\caption{(a) Spatial distribution of the rate of axial extension rate $\frac{\partial V_x}{\partial x}$. The full lines indicate the location of the die walls.(b) Sample profiles of the rate of axial extension at several axial locations downstream: diamonds - $x= -3 ~mm$, circles - $x=0~mm$, squares-$x=0.1~mm$, stars - $x=0.3~mm$, triangles - $x=0.6~mm$ . The data refers to experiment $1$, Table \ref{tab_tab2}. } 
\label{gradmapLDPE}
\end{center}
\end{figure*}

Similar measurements of the spatial distribution of the velocity gradients have been performed for the stable (LDPE) case, Fig. \ref{gradmapLDPE}.

Although no significant difference between the stable and unstable flow fields could be found (Figs. \ref{flowmapLLDPE}, 
\ref{flowmapLDPE}), subtle differences in the fields of velocity gradients are visible.
A first important difference can be noted at the die exit: in the stable case (Fig. \ref{gradmapLDPE}(b)) the absolute values of the velocity gradients near the boundaries and in the bulk are much closer than in the unstable case (Fig. \ref{gradmapLLDPE} (b)). 
A second difference between the two fields of axial velocity gradients can be noticed in the bulk of the flow: the local minima and the maximum are less pronounced in the stable case.

Though the description of the flow kinematics is instructive and certainly needed while exploring the sharkskin, a deeper insight into the phenomenon needs a good understanding of the rheological behavior of the material as well. To this is dedicated the next section.

\subsection{Extensional rheological measurements}\label{subsec:rheology}

As already noted by others \cite{dennreview, denn3, arda}, a complete picture of the sharkskin phenomenon requires the knowledge of both kinematic and dynamic properties of the flow. Dynamical properties of the flow are directly related to the rheological behavior of the material under investigation.
The extensional rheological properties  have been investigated for both LLDPE and LDPE.  Of particlular interest for the investigation of the sharkskin instability is the melt strength of the material defined as the stress $\sigma_b$ at which the sample breaks during the extensional test. In the case a sample does not break it can only be said that the melt strength is higher than the maximum stress $ \sigma_{m} $ measured.     The melt strength is usually measured using the Rheotens apparatus. The M\"{u}nstedt type extensional rheometer employed in this study is able, however, to deliver more accurate results than the Rheotens device as it operates under isothermal conditions. The reproducibility of the stress strain curves in the stressing experiments (constant rate of extension, $\dot \epsilon$) for the LLDPE and the LDPE is better than $5~\%$.

\begin{figure*}
\begin{center}
\includegraphics[width=12cm]{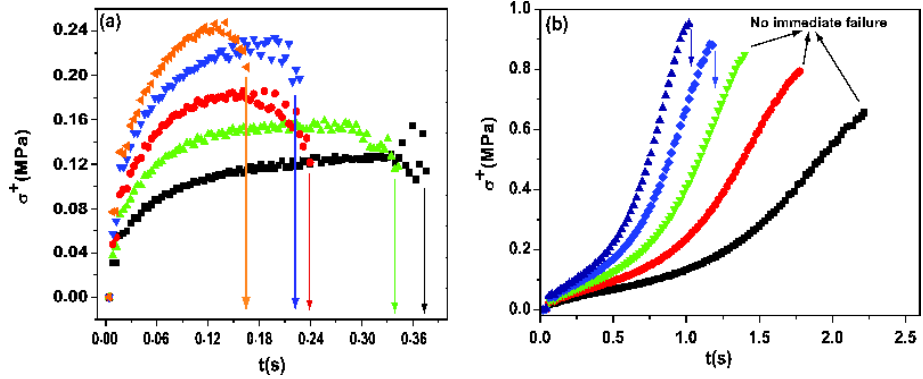}
\caption{(a)Transient stress measurements for the LLDPE at different rates of elongation and $T=220^{\circ} C$: squares -$\dot {\epsilon}= 6.7 s^{-1}$, up triangles - $\dot {\epsilon}= 8.7 s^{-1}$, circles - $\dot {\epsilon}= 10.6 s^{-1}$, down triangles - $\dot {\epsilon}= 12.5 s^{-1}$, left triangles - $\dot {\epsilon}= 14.5 s^{-1}$. (b) Transient stress measurements for the LDPE at different rates of elongation and $T=135^{\circ} C$: squares -$\dot {\epsilon}= 1.4 s^{-1}$, circles - $\dot {\epsilon}= 1.8 s^{-1}$, down triangles - $\dot {\epsilon}= 2.2 s^{-1}$, rhumbs - $\dot {\epsilon}= 2.6 s^{-1}$, up triangles - $\dot {\epsilon}= 3 s^{-1}$. The vertical arrows indicate the physical rupture of the sample.}\label{extrheo}
\end{center}
\end{figure*}

Measurements of the transient extensional stresses for the LLDPE at $T=220^{\circ} C$ are presented in Fig. \ref{extrheo} (a).
The extensional data has been shifted from $T=135^{\circ} C$ using the time temperature superposition principle, in order to extend the range of rates of extension closer to the rates achieved during the extrusion experiment (experiment 7 in Table \ref{tab_tab2}).  This way, a correlation between the flow kinematics discussed in Sec. \ref{subsec:kinematics} and the rheological properties of  the material is possible. For each of the rates of extension investigated, no significant strain hardening effect has been observed. It can also be noted that a steady state of extension (a plateau) seems to be reached for each of the rates of deformation investigated prior to the physical rupture of the sample.

Similar measurements were carried out with the LDPE Fig. \ref{extrheo} (b). There exist several significant differences between the LDPE and the LLDPE extensional data. Firstly, in contrast to the LLDPE, only the samples at the two highest elongational rates broke during elongation, Fig. \ref{extrheo} (b).  In the other cases the samples could be stretched to the maximum strain of $ 3.2 $ and failed shortly afterwards due to interfacial effects. The points at which the specimens broke are indicated by arrows. Secondly, unlike in the experiments with LLDPE, no steady state seems to be reached in the experiments with LDPE. 
Thirdly, the extensional tests with LDPE indicate a significant strain hardening.

\begin{figure*}
\begin{center}
\includegraphics[width=8cm]{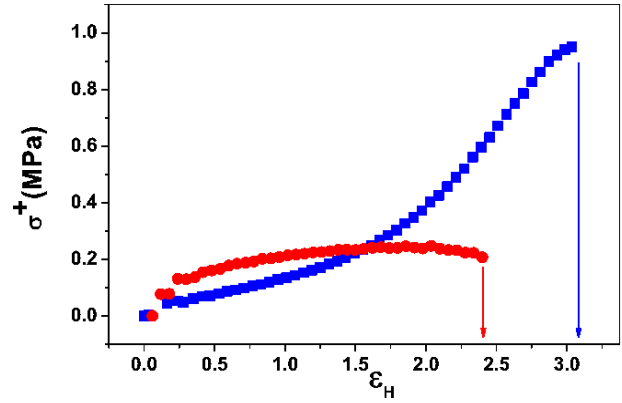}
\caption{Comparison of the stress-strain curves for LDPE and LLDPE: squares - LDPE  at $\dot{\epsilon}=3.0 s^{-1}$ and $T= 135^ \circ C$, circles - LLDPE at $\dot{\epsilon}= 14.5 s^{-1}$ and $T= 220^ \circ C$. The vertical arrows indicate the physical rupture of the samples.}\label{extrheoBOTH}
\end{center}
\end{figure*}
A comparison of two typical stress-strain curves of the two materials is presented in Fig. \ref{extrheoBOTH}. The shape of the two curves differs completely which indicates that the failure mechanisms are, most probably, different for the two materials, too. 
Finally, we note that the stress at break is much higher in the case of LDPE than in the case of the LLDPE which simply means that, at a given flow rate through the die, the LDPE can sustain larger stresses prior to failure.

\begin{figure*}
\begin{center}
\includegraphics[width=7cm]{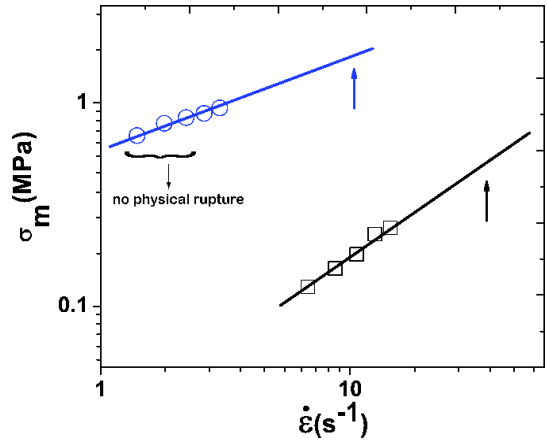}
\caption{Comparison of the maximum stresses obtained for LDPE and LLDPE: squares - LLDPE , $T= 220^ \circ C$, circles - LDPE $T= 135^ \circ C$. The full lines are power law fits. The LLDPE data has been shifted from $T= 135^ \circ C$ using the time-temperature superposition principle. The vertical arrows indicate the maximal rates of extension reached near the boundaries during the extrusion experiments.}\label{meltstrength}
\end{center}
\end{figure*}

Fig. \ref{meltstrength} shows a comparison of the highest stresses reached during the extensional tests as a function of the rate of extension $\dot \epsilon$. Each point represents the average of at least five measurements. As seen before, the maximum stresses obtained increase with growing shear rates. LDPE displays maximum stresses that are three to four times higher than the values for LLDPE.
For both materials the dependence of the maximal stresses on the rate of extension can be fitted by a power law (the full lines in Fig. \ref{meltstrength}) and this will be used in the next section for assessments of the distribution of the extensional stresses in the flow.

\subsection{Measurements of the stress distribution in the flow} \label{subsec:stressdistribution}
\subsubsection{Distribution of extensional stresses} \label{subsec:extensionalstress}

To understand the distribution of both extensional and shear stresses in the flow (particularly near the die exit), we connect the information on the flow kinematics presented in Sec. \ref{subsec:kinematics} with the rheological properties of the materials discussed in Sec. \ref{subsec:rheology}.
Because the extensional stress is a function of both time and rate of extension, $\sigma_{xx}=\sigma_{xx}(t, \dot{\epsilon})$,  a rigorous calculation of the distribution of the extensional stresses is a nontrivial task. The time dependence mentioned above requires tracking material elements on a Lagrangian trajectory which is a technically non-trivial task. In our case, however, a reasonable simplification can be made. For this purpose, we show that the characteristic flow time through the die $\tau_{char} \propto \frac{\Delta l}{ V_x^{av}}$ needed for the material to readjust itself from a non-slip boundary condition (within the die) to a free boundary condition (outside the die) is comparable to the characteristic time $t_R$ at which physical rupture of the samples occurs during extensional measurements. Here $\Delta l$ denotes the characteristic length over which the material readjusts to the new boundary conditions and $V_x^{av}$ is the average (along y axis) velocity at the die exit ($x=0$).
\begin{table*}
\begin{center}
  \begin{tabular}{@{} |c|c|c|c|c|c| @{}}
    \hline
   \textbf{Material} & $\mathbf{Experiment}$ & $\mathbf{\Delta l (~mm)}$ & $\mathbf {V_x^{av} (mm~s^{-1})}$ & $\mathbf{\tau_{char} (s)}$ &$\mathbf {t_R (s)}$ \\    \hline \hline
      LLDPE & $7$ & $1$ & $7.03$ & $0.14$ &$0.17$ \\    \hline \hline
    LDPE& $1$ & $1.8$ & $2.19$ & $0.82$ &$1$ \\    \hline 
   \end{tabular}
   \end{center}
 \caption{Comparison of the characteristic flow times with the failure times.} \label{tab_tab3}
\end{table*}

Estimates for the characteristic length $\Delta l$ and the average velocity $V_x^{av}$ at the die exit can be obtained from Fig. \ref{flowmapLLDPE} for the experiment with LLDPE and from Fig. \ref{flowmapLDPE} for the experiment with LDPE.
Values for the times $t_R$ at which physical rupture of the samples occurs during extensional measurements can be obtained from Fig. \ref{extrheoBOTH}.

In Table \ref{tab_tab3} we present a comparison of such estimates for the two materials investigated. Indeed, one can clearly note that $\tau_{char}$ and $t_R$ have similar values for both materials. A similar conclusion can be reached if one estimates the characteristic flow times using average values of the axial velocity gradients, $\tau_{char} \propto \left ( \left  | \frac{\partial V_x}{\partial x} \right |_{av} \right )^{-1}$ where the average is taken along the $y$ direction. 

This allows a simpler calculation of the distribution of the extensional stresses in the flow by considering only the rate dependence of the maximal stress attained during extensional tests, $\sigma_m=\sigma_m(\dot \epsilon)$ and using as rate of extension $\dot \epsilon$ the measured two dimensional velocity gradient, $\frac{\partial V_x}{\partial x}$, Figs. \ref{gradmapLLDPE}, \ref{gradmapLDPE}. 

Unfortunately, due to a technical limitation of the M\"{u}nstedt rheometer, the maximal rates of extension attainable during the rheological measurements presented in Fig. \ref{meltstrength} were two to three times smaller than the maximal values of the axial velocity gradients measured around the die exit in the proximity of the die walls. For the characterization of the LLDPE (which is of particular importance, as it shows sharkskin) this mismatch has been reduced by shifting measurements of the transient extensional viscosity from $T=135 ^{\circ}~C$ to $T=220 ^{\circ}~C$ using the time-temperature superposition principle.  Because of this limitation we have used for the calculation of the distribution of extensional stresses the power law fits presented as full lines in Fig. \ref{meltstrength} and extrapolate these dependencies to the appropriate range of velocity gradients indicated by the vertical arrows in Fig. \ref{meltstrength}. Though we agree that this might not be the most rigorous procedure to calculate the distribution of the extensional stresses in the flow (we do not know whether the power law dependencies fitted in Fig. \ref{meltstrength} are preserved during the extrapolation at higher rates of extension), we still believe that this approach is a reasonable compromise (and, to our understanding, the best could be done in this context) and can still shed some light on the phenomenology of the sharkskin formation.  We also note that, in fact, the extrapolation is used only in a limited region of the flow, near the boundaries.   

A second shortcoming of this procedure is that the flow kinematics during extensional rheological tests and the extrusion experiments is different: whereas during the rheological tests we dealt with an uniaxial extension, the flow through the die combines both shear and planar extension.  We therefore conclude that this approach can only provide a semi-quantitative picture of the phenomena.

\begin{figure*}
\begin{center}
\includegraphics[width=14cm]{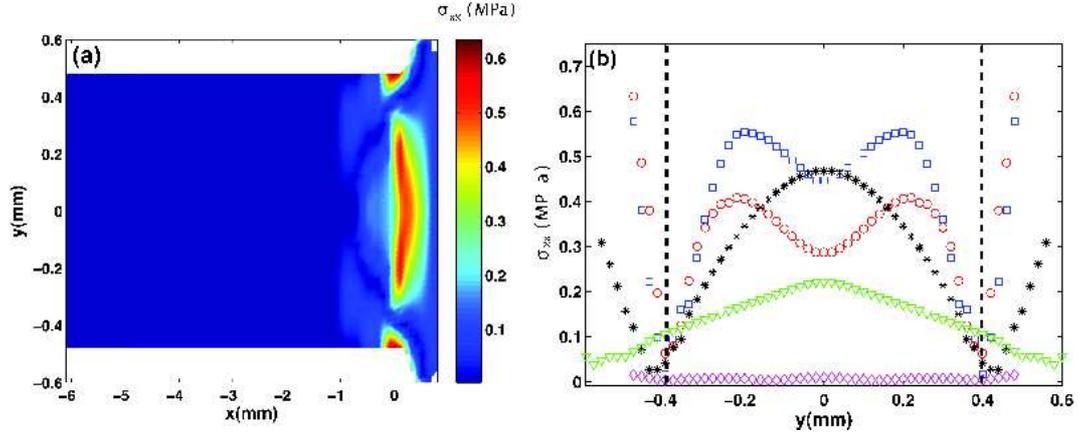}
\caption{(a) Spatial distribution of the extensional stresses $\sigma_{xx}$ for LLDPE above the onset of sharkskin. (b) Profiles of the extensional stresses across the flow direction at several axial positions: diamonds - $x= -3 ~mm$, circles - $x=0~mm$, squares-$x=0.1~mm$, stars - $x=0.3~mm$, triangles - $x=0.6~mm$ . The data refer to experiment $7$, Table \ref{tab_tab2}. The vertical dotted lines indicate the locations of the edges of the stress boundary layer (see text).}\label{extensionalstressesLLDPE}
\end{center}
\end{figure*}

Calculation of the distribution of extensional stresses is presented for LLDPE in Fig. \ref{extensionalstressesLLDPE}. As the flow within the die is a shear flow, no extensional stresses are present for $x<0$ (diamonds Fig.  \ref{extensionalstressesLLDPE} (b)).
The inhomogeneity in the distribution of the axial velocity gradients visible in Fig. \ref{gradmapLLDPE} becomes even more pronounced: a nearly two-fold stress imbalance between the boundary and the bulk of the flow can be observed at the die exit, circles Fig. \ref{extensionalstressesLLDPE} (b). Corresponding to the location (along $y$ axis) where the axial gradients of the axial velocity component change sign Fig. \ref{gradmapLLDPE} (b) the extensional stresses pass through a minimum. We interpret this fact as the emergence of a boundary layer for the extensional stresses within which the material elements are significantly stretched along the flow direction (see the vertical dotted lines in Fig. \ref{extensionalstressesLLDPE} (b)). 
It is very important to note that the extrapolated value of the melt strength ($\sigma_m=0.5~MPa$ at $\dot{\epsilon}=40~s^{-1}$) is comparable to the average stress imbalance between the boundary layer and bulk: $\sigma_{xx}^{boundary}-\sigma_{xx}^+{bulk} = 0.37~MPa$. This suggests that the material located within the boundary layer fails during the extrusion process and sharkskin is observed. 

Over a distance as small as $0.1~mm$ from the die exit, the stress imbalance visible at $x=0~mm$ diminishes rapidly and the stresses in the boundary layer are now comparable to stresses in the bulk, squares Fig.  \ref{extensionalstressesLLDPE} (b).  Simultaneous with this redistribution of extensional stresses between the boundaries and the bulk, the sharkskin instability emerges. Further downstream, the stress accumulation in the boundaries disappears  gradually (stars and triangles, Fig.  \ref{extensionalstressesLLDPE} (b)).

\begin{figure*}
\begin{center}
\includegraphics[width=14cm, height=5.5cm]{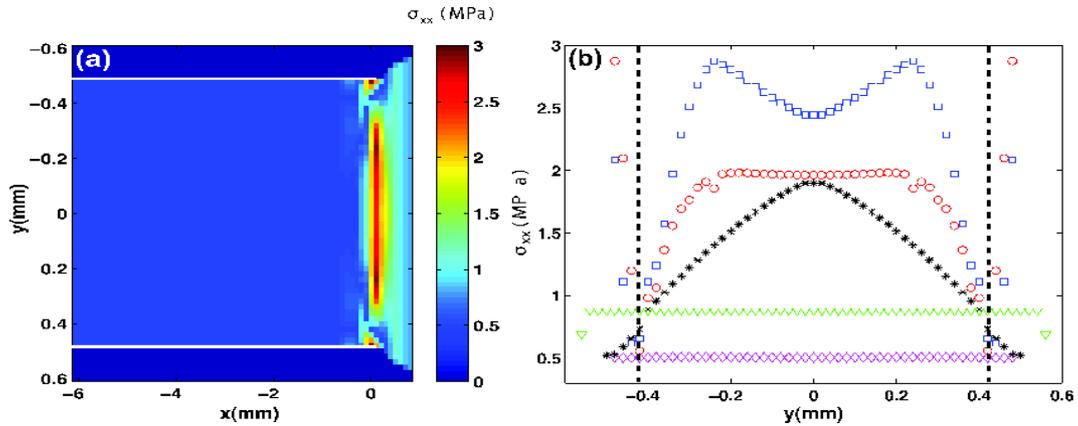}
\caption{(a) Spatial distribution of the extensional stresses $\sigma_{xx}$ for LDPE . The horizontal white lines indicate the position of the die wall.(b) Profiles of the extensional stresses across the flow direction at several axial positions: diamonds - $x= -3 ~mm$, circles - $x=0~mm$, squares-$x=0.1~mm$, stars - $x=0.3~mm$, triangles - $x=0.6~mm$. The data refer to experiment $1$, Table \ref{tab_tab2}. The vertical dotted lines indicate the locations of the edges of the stress boundary layer (see text).}\label{extensionalstressesLDPE}
\end{center}
\end{figure*}

Measurements of the distribution of extensional stresses are presented for LDPE in Fig. \ref{extensionalstressesLDPE}. Clear differences with respect to the stress distribution for LLDPE (above the onset of sharkskin) can be observed. First, the stress distribution around the die exit is in this case more homogeneous than in the case of LLDPE. Indeed, the stress values in the boundaries are in this case only about $30 \%$ higher than in the bulk. We interpret this as a more stable stress distribution along the $y$ axis. A second notable difference is related to the extent of the boundary layer: in the case of the LDPE the bulk region is wider and the boundary layer thinner than in the (unstable) LLDPE case. This is, in our opinion, a stabilizing factor as well. 
The extrapolated value of the melt strength ($\sigma_m=1.7~MPa$ at $\dot{\epsilon}=15~s^{-1}$) is nearly two times larger than the average stress imbalance between the boundary layer and bulk: $\sigma^+_{boundary}-\sigma^+_{bulk} = 1~MPa$. This suggests that the material can sustain the extensional stress imbalance without failing which corroborates with the absence of the sharkskin instability. 
A third difference between the stress distributions for the LLDPE and the LDPE experiments relates to the axial extent of the stress distributions, Figs. \ref{extensionalstressesLLDPE}, \ref{extensionalstressesLDPE} (a).
Whereas in the case of the LLDPE the local stress maxima visible in the bulk region of the flow extend over nearly two millimeters along the flow direction ($x$ axis) Fig. \ref{extensionalstressesLLDPE} (a), in the stable case (LDPE) they extend over less than a millimeter, Fig. \ref{extensionalstressesLDPE} (a). This can be qualitatively understood as follows. Due to a smaller stress imbalance between the boundary and the bulk, the material readjusts itself from a non-slip boundary condition to a plug like flow faster (over a shorter axial distance) in the case of LDPE flow but it needs more time in the case of the LLDPE experiments.

Finally we note that the simplified way of calculating the stress distribution in the flow (by neglecting the time dependence of the tensile stresses) proposed in the beginning of this section does not alter the validity of our conclusions, as these conclusions are stated in terms of stress differences (between boundary and bulk) rather than in terms of stresses.

\subsubsection{Distribution of shear stresses} \label{subsec:shearstress}
Different flow regimes during the extrusion of polymer melts through dies have been traditionally characterized in terms of the wall shear stress (calculated using the pressure drop) and the apparent shear rates \cite{denn3, dennreview}, as illustrated in the sketch presented in Fig. \ref{fig_phasediagram}. On the other hand, it has been demonstrated in Sec. \ref{subsec:extensionalstress} that a main difference between the unstable and stable flows is actually related to the distribution of extensional stresses around the die exit, more precisely to the stress balance between the boundaries and the bulk of the flow. 

In this context, a natural question arrises: do the shear stresses play any role in the emergence of the sharkskin instability and furthermore, is the wall shear stress a true control parameter or just a formal one?

\begin{figure*}
\begin{center}
\includegraphics[width=14cm]{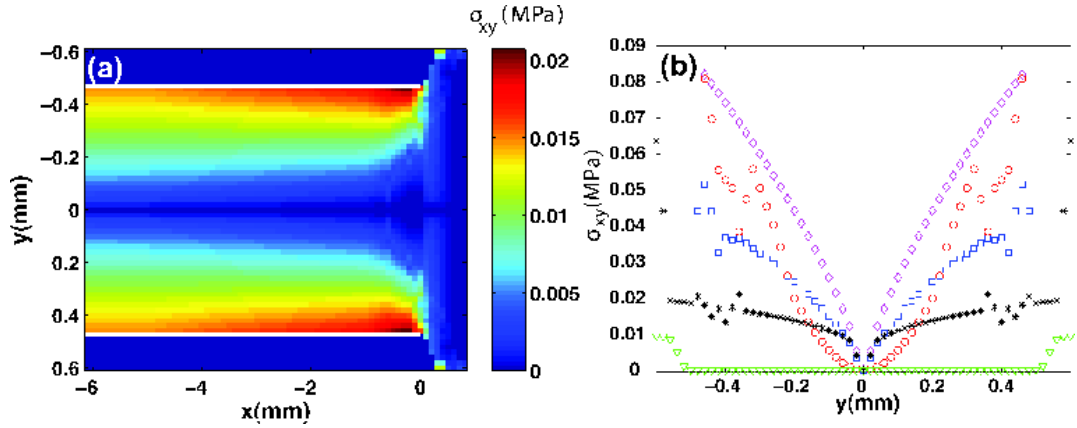}
\caption{(a) Spatial distribution of the shear stresses $\sigma_{xy}$ for the LLDPE above the onset of sharkskin. The horizontal white lines indicate the position of the die wall.(b) Profiles of the shear stresses across the flow direction at several axial positions: diamonds - $x= -3 ~mm$,  circles - $x=0~mm$, squares-$x=0.1~mm$, stars - $x=0.3~mm$, triangles - $x=0.6~mm$. The data refer to experiment $7$, Table \ref{tab_tab2}.}\label{shearstressesLLDPE}
\end{center}
\end{figure*}

\begin{figure*}
\centering
\begin{center}
\includegraphics[width=14cm]{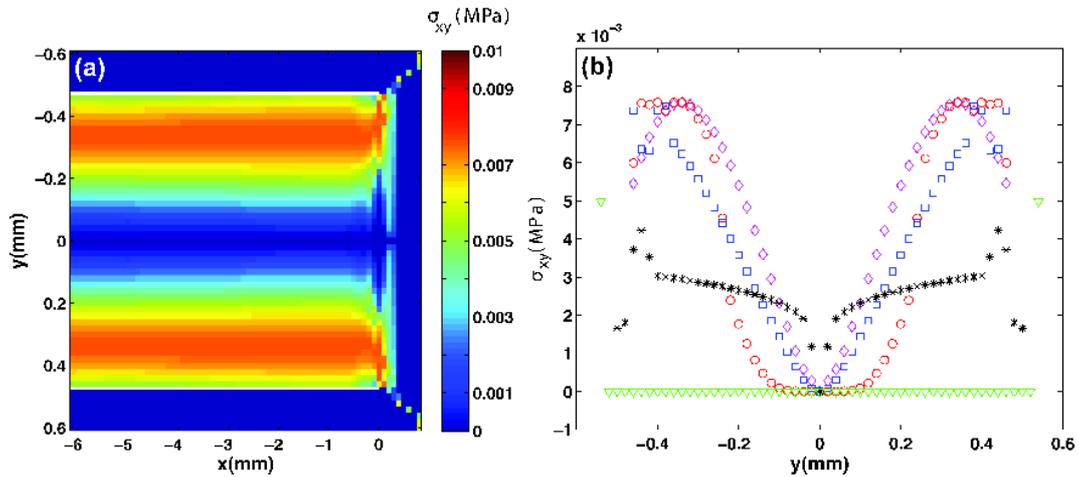}
\caption{(a) Spatial distribution of the shear stresses $\sigma_{xy}$ for the LDPE. The horizontal white lines indicate the position of the die wall.(b) Profiles of the shear stresses across the flow direction at several axial positions: circles - $x=0~mm$, squares-$x=0.1~mm$, stars - $x=0.3~mm$, triangles - $x=0.6~mm$ .The data refers to experiment $1$, Table \ref{tab_tab2}.}\label{shearstressesLDPE}
\end{center}
\end{figure*}

To address this question, we present in the following a comparative study of the distributions of shear stresses in both unstable (LLDPE) and stable (LDPE) cases. Two dimensional maps of the shear stresses are obtained using a procedure similar to that described in Sec. \ref{subsec:extensionalstress}.  First, shear rheological measurements have been carried out for both materials and the dependence of the shear stresses $\sigma_{xy}$ on the rate of shear $\dot \gamma$ has been obtained (data not shown here). Secondly, the shear flow curves have been fitted with the Carreau-Yassuda model and the distribution of the shear stresses in the flow has been calculated using the fit function and interpreting the transversal velocity gradients $\frac{\partial V_x}{\partial y}$ as rates of shear.
Measurements of the distribution of the shear stresses for LLDPE and LDPE are shown in Figs. \ref{shearstressesLLDPE},  \ref{shearstressesLDPE}, respectively.

 For both materials investigated, the flow within the die ($x<0$) is consistent with a fully developed Poiseuille flow characterized by shear stresses monotonically decreasing from a maximal value near the boundaries ($y=\pm 0.5$) to zero at the center line ($y=0$).
Around the die exit the shear stresses quickly decay to zero for both materials which corresponds to a transition to a plug-like flow. For both the unstable (LLDPE) and stable (LDPE) cases the maximal shear stresses attained within the die near the walls are significantly smaller than the extensional stresses which suggests that the emergence of the sharkskin instability is, most probably, unrelated to the shear stresses. 

Therefore one can conclude that the wall shear stress can be used only as a formal control parameter in studying different flow regimes. The fact that diagrams such as the one sketched in Fig. \ref{fig_phasediagram} still provide useful information regarding the stability of the extrusion flow is that the extensional stresses are increasing with the pressure drop as well. 

\section{Conclusions} \label{Conclusions}

A systematic experimental study of the sharkskin instability has been presented. The flows of two polymer melts (namely LLDPE, which shows sharkskin for apparent shear rates larger than an onset value and LDPE, which shows no sharkskin over a broad range of apparent shear rates)  through a slit die have been accurately characterized using LDA. 

The central idea of the present study is to analyze separately the flow kinematics and the rheological properties of the materials (in both extension and shear) and based on this to obtain the distribution of extensional and shear stresses in the flow by connecting the kinematic flow properties with the rheological properties.  

In Sec. \ref{subsec:kinematics} we have compared the kinematics of unstable and stable extrusion flows. The flow fields are qualitatively similar in both cases: as the materials exit the die the flows accelerate near the boundaries and decelerate around the center line, Figs. \ref{flowmapLLDPE}, \ref{flowmapLDPE}. 

This apparent similarity of the unstable and stable flow fields prompted us to perform a deeper analysis of the flow kinematics by analyzing the fields of the axial gradients of the axial velocity component, $\frac{\partial V_x}{\partial x}$. Unlike the flow fields, the fields of gradients for the unstable and stable cases do show several notable differences. In the unstable case there exists a larger imbalance of the velocity gradients between the boundary region and the bulk region than in the stable case, Figs. \ref{gradmapLLDPE}, \ref{gradmapLDPE}. This indicates that corresponding to the unstable case the surface layers of the material are subjected to higher accelerations (or equivalently rates of extension)  than  in the stable case. In the unstable case one can also note that at the die exit the transversal distribution of the axial velocity gradients is more inhomogeneous than in the stable case. Indeed, whereas the transversal profiles of velocity gradients at the die exit exhibit pronounced local maxima for the LLDPE (circles, Figs. \ref{gradmapLLDPE}(b)), the transversal profiles for the LDPE exhibit less pronounced or no local maxima (circles and squares, Figs. \ref{gradmapLDPE}(b)).  

Sec. \ref{subsec:rheology} presents a comparative study of the rheological properties of LLDPE and LDPE in uniaxial extension. A major difficulty is related to matching the range of the rates of extension accessible by the M\"{u}nstedt rheometer with the range of velocity gradients measured by LDA in the extrusion flows. This is particularly important for the case of LLDPE(where the sharkskin instability is observed). As the maximal rate of deformation ($\dot \epsilon_{max}=3~s^{-1}$) achievable during the rheometric tests could not be increased, the only way to overcome this difficulty was using the time-temperature superposition principle to shift the extensional data acquired at $T= 135^ \circ C$ to $T= 220^ \circ C$ (which is the temperature at which the flow data was acquired - experiment 7, Table \ref{tab_tab2}).  
Thus, the maximal rate of deformation during the extensional tests becomes $\dot \epsilon_{max}=14.5~s^{-1}$ which compares better to the average values of the axial velocity gradients at the die exit (roughly $20~s^{-1}$ in the bulk and up to $40~s^{-1}$). 

Measurements of the transient tensile stresses reveal a significantly different rheological behavior of the two materials, Figs. \ref{extrheo}, \ref{extrheoBOTH}. The tensile stress measurements for LLDPE show no significant strain hardening behavior and a steady state seems to be reached prior to the physical rupture of the sample (which occurred for each of the rates of deformation investigated), Fig. \ref{extrheo} (a). The tensile stress measurements for LDPE show a significant strain hardening behavior and for each of the rates of extension observed, no steady state has been reached, Fig. \ref{extrheo} (b). Corresponding to the highest rates of the deformation explored, the LDPE samples break at a higher Hencky strain and the terminal stress at the break point is about $4.6$ times larger than in the case of LLDPE, Fig. \ref{extrheoBOTH}. 
The main conclusion of the rheological investigation is that, at a given deformation (Hencky strain), the LDPE is able to sustain much higher tensile stresses prior to failure. 

Sec. \ref{subsec:stressdistribution} connects the results on the flow kinematics presented in Sec. \ref{subsec:kinematics} with the rheological characterization presented in Sec. \ref{subsec:rheology} in order to characterize the distribution of both tensile and shear stresses during the extrusion flows of the two materials. Spatial distributions of the tensile stresses have been measured using the rate dependence of the maximal stresses reached during tensile tests Fig. \ref{meltstrength} and the spatial distribution of the axial gradients of the axial velocity component, Figs. \ref{gradmapLLDPE}, \ref{gradmapLDPE}. The inhomogeneity in the transversal distribution of the velocity gradients observed near the die exit during the experiments with the LLDPE translates into an even stronger inhomogeneity in the transversal distribution of the extensional stresses, Fig. \ref{extensionalstressesLLDPE} (b). This inhomogeneity is less pronounced or absent in the case of LDPE, Fig. \ref{extensionalstressesLDPE} (b), suggesting that this transversal stress distribution relates to a more stable configuration. 

Of particular relevance is the emergence of a boundary layer for the tensile stresses, Figs.  \ref{extensionalstressesLLDPE} (b), \ref{extensionalstressesLDPE}(b). Corresponding to the edge of the boundary layer the profiles of the axial gradients measured outside of the die intersect (Figs. \ref{gradmapLLDPE} (b), \ref{gradmapLDPE} (b)) and the corresponding tensile stresses pass through a minimum, Figs.  \ref{extensionalstressesLLDPE} (b), \ref{extensionalstressesLDPE}(b). In the unstable case (LLDPE) the tensile stress in the boundary layer is roughly twice larger than the tensile stress in the bulk at the die exit. The stress imbalance between boundary and bulk is in this case practically equal to the melt strength of the material Fig. \ref{meltstrength}. As a consequence, the materials fails, the sharkskin instability emerges and the stress imbalance is released right after the die exit (squares, Fig.  \ref{extensionalstressesLLDPE} (b)). 

In the stable case (LDPE) the stresses in the boundary are only about $30 \%$ larger than in the bulk and this imbalance is smaller than the melt strength. Consequently the material does not fail near the die exit and no sharkskin instability is observed.
Another important difference between the unstable and the stable case relates to the extent of the stress boundary layer: in the unstable case the boundary layer is wider. Bearing in mind that the integral of the tensile stress over the boundary layer $\int_{BL} \sigma_{xx}(y)dy$ represents a surface energy density, one can conclude that in the unstable case more energy is accumulated in the superficial layers than in the stable case.
Measurements of the distribution of the shear stresses at the die exit and outside the die for both materials reveal a quick transition from a Poisseuile flow to a plug flow characterized by negligible values of the shear stresses. Therefore a role of the shear stresses in the emergence of the sharkskin instability is ruled out.

\section*{Acknowledgments}

T. I. B. gratefully acknowledges the financial support from the German Research Foundation, grant MU 1336/6-4.
We are grateful to J. Resch and U. Kessner for their support with the time-temperature shift of the LLDPE elongational data. 
Furthermore, we thank the referees for their valuable comments and suggestions. 

\bibliographystyle{elsarticle-harv}

\end{document}